\documentclass[conference]{IEEEtran}
\IEEEoverridecommandlockouts
\usepackage{cite}
\usepackage{amsmath,amssymb,amsfonts}
\usepackage{algorithmic}
\usepackage{graphicx}
\usepackage{textcomp}
\usepackage{xcolor}
\def\BibTeX{{\rm B\kern-.05em{\sc i\kern-.025em b}\kern-.08em
    T\kern-.1667em\lower.7ex\hbox{E}\kern-.125emX}}
\begin{document}

\title{Synchronization in Digital Twins for Industrial Control Systems\\
}

\author{\IEEEauthorblockN{Fatemeh Akbarian\textsuperscript{1}, Emma Fitzgerald\textsuperscript{1,2}, Maria Kihl\textsuperscript{1}}
\IEEEauthorblockA{\textit{\textsuperscript{1}Department of Electrical and Information Technology, Lund University, Sweden}}
\IEEEauthorblockA{\textit{\textsuperscript{2}Institute of Telecommunications, Warsaw University of Technology, Poland}} \\
}

\maketitle

\begin{abstract}
Digital twins, which are a new concept in industrial control systems (ICS), play a key role in realizing the vision of a smart factory, and they can have different effective use cases. With digital twins, we have virtual replicas of physical systems so that they precisely mirror the internal behavior of the physical systems. Hence, synchronization is necessary to keep the states of digital twins in sync with those of their physical counterparts. Otherwise, their behavior may be different from each other, and it can lead to wrong decisions about the system that can have catastrophic consequences. In this paper, we propose three different architectures for digital twins, and then by investigating their ability to follow the physical system’s behavior, we will determine the best architecture, whose output has the lowest error compared with the physical system's output.
\end{abstract}

\begin{IEEEkeywords}
Digital Twin, Smart factory, Synchronization
\end{IEEEkeywords}

\section{Introduction}
Industrial control system (ICS) is a general term that encompasses several types of control systems, including supervisory control and data acquisition (SCADA) systems, distributed control systems (DCS), and other control system configurations. 
Along with advanced technologies in ICS, such as big data, Internet-of-Things (IoT), cloud computing, etc, the current manufacturing industry is going through an unprecedented dramatic change in the last decade, and smart manufacturing has attracted much attention\cite{cheng2018industrial}. 

Digital Twin is one of the main concepts associated with smart manufactoring, and it opens up new possibilities in terms of monitoring, simulating, optimizing and predicting the state of cyber-physical systems (CPSs). A digital twin duplicates the physical model for remote monitoring, viewing, and controlling based on the digital format, and its main goal is to closely follow products during production (the physical twin) and simulate the process to adjust the production with the results of these simulations\cite{farsi2020digital}. 

The concept of digital twins has its origins in NASA’s Apollo program, as a twin of a spacecraft was built for two purposes, viz., (i) training before the mission and (ii) supporting the mission by mirroring flight conditions based on data coming from the spacecraft in operation\cite{glaessgen2012digital,grieves2014digital}. 
Motivated by the need to utilize machine or process data for the purpose of prognostics, digital twins were proposed for production systems in the cloud that simulate the conditions of their physical counterparts based on physical models\cite{eckhart2019digital}. 

In smart factories, the digital twin is located in cloud resources and includes simulation of all components and units of the physical part, so it allows advanced simulations of a complete manufacturing system.  These simulations in the digital twin should be done in real-time to enable it to track the physical part's behavior. Furthermore, the virtual part should be able to imitate its physical counterpart accurately. Hence, synchronization between the digital and physical parts is crucial. 

So far, different architectures, as well as use cases, have been proposed for digital twins. However, most of these works have missed the synchronization between the digital and physical parts. For example, the authors in\cite{eckhart2018towards} propose a  framework for digital twins that can be automatically generated from the specification and run independently from the physical environment. In\cite{jain2019digital}, a digital twin has been developed that estimates the characteristics of a photovoltaic energy conversion system. This digital twin is created by using the mathematical model of the system and it does not follow the physical system’s behavior but rather estimates the correct behavior of the system. Hence, although it is useful for fault detection in the system, there is no synchronization between the digital and physical twins which is the most important part of the definition of digital twin.

There are a few works that have suggested replication and synchronization but only for some limited cases. To the best of our knowledge, \cite{eckhart2018specification} is the only work that has suggested a method for synchronization. In\cite{eckhart2018specification}, a passive state replication approach has been proposed. In this method, the inputs of the physical system that constitutes a stimulus should be fed to the digital twin. For example, a setpoint that a user chooses for the system through the HMI (Human-Machine Interface) is a kind of data that should be replicated in the digital twin. However, using passive monitoring in this approach may result in missing some important data related to the physical system’s states. Also, this method does not allow the digital twin to follow the physical system continuously; for instance when there are unexpected changes in the system like faults, the digital twin is not able to imitate these changes.

Motivated by these challenges, we make the following novel contributions in this paper:
\begin{itemize}
\item We propose three different architectures for digital twins.
\item We evaluate the ability of these proposed architectures to mirror their physical counterpart’s behavior.
\item We demonstrate which architecture from these proposed architectures has the best performance in keeping the digital twin in sync with the physical system.
\end{itemize}

\section{Targeted System}
\begin{figure}[htp]
    \centering
    \includegraphics[width=6cm]{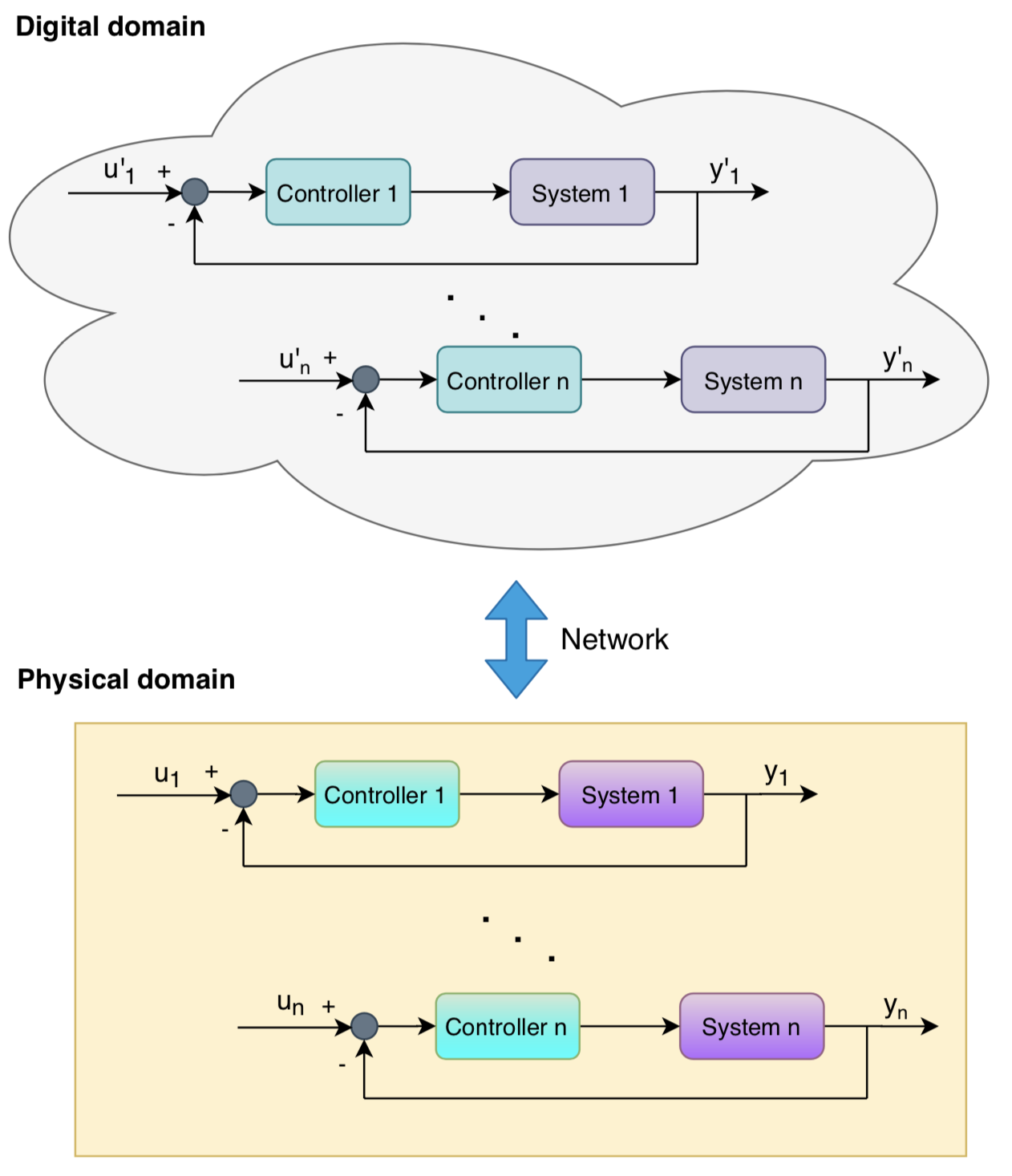}
    \caption{Targeted system overview}
    \label{fig:fig1}
\end{figure}

The targeted system, which concerns industrial control systems, is illustrated in Fig.~\ref{fig:fig1}. In this figure, the physical domain is inside a factory, which consists of several different systems, and each of these systems needs to be controlled by their local controller. The output and input of the \(i\)th system in the physical domain and their counterparts in the digital domain have been denoted respectively by \(y_{i}\), \(u_{i}\), \(y_{i}'\), and \(u_{i}'\). With digital twins, we have simulated models of each real system in the physical part. These models can be defined by using data-driven methods or they can be created based on laws of physics. However, since most of the machines in a factory are complex systems, utilizing data-driven methods like system identification algorithms can be the best choice. In this approach, we feed some different input signals to the system and record output signals. Then, based on these inputs, their outputs, and system identification algorithms, we can find the best model for the system.

As can be seen in Fig.~\ref{fig:fig1}, digital twins are located in a cloud and they interact with the physical domain through the network. As was said before, the main goal of digital twins is following the physical systems' behavior. In order to fulfill this, synchronization is necessary to keep the states of digital twins consistent with those of their physical counterparts. Therefore, the challenging issue is how the architecture of digital twins should be constructed and which signals from the physical domain should be sent to the digital domain.
\section{Proposed Solution}
In order to make digital twins able to follow the physical systems continuously so that they can imitate physical systems’ behavior even when there are unexpected changes in the system, we propose three different architectures for implementing digital twins. In each of these architectures, we \\
\begin{figure}[htp]
    \centering
    \includegraphics[width=7.5cm, height=3cm]{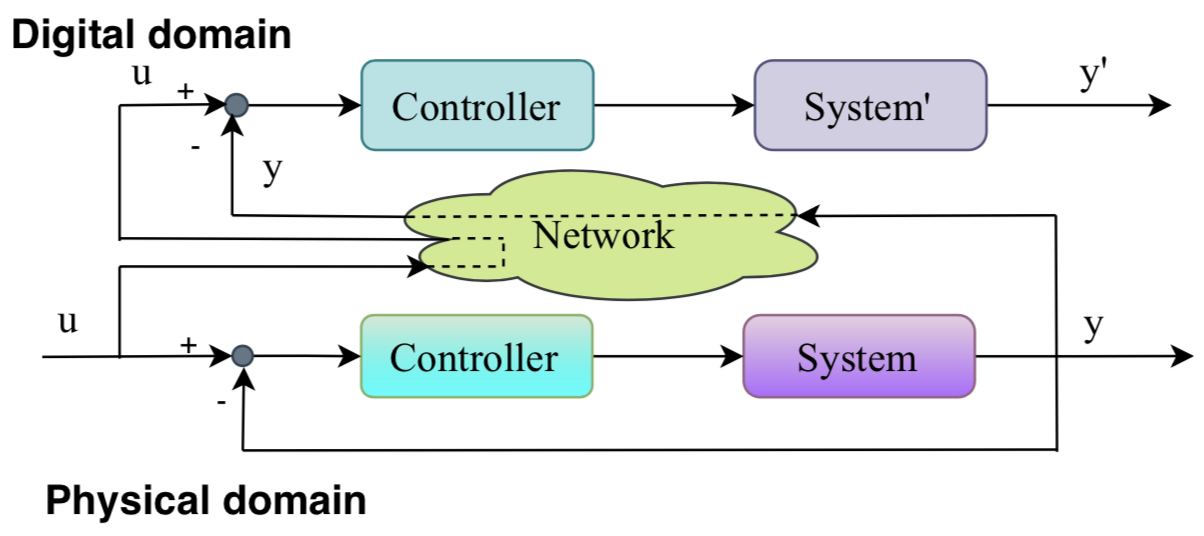}
    \caption{Architecture I}
    \label{fig:fig2}
\end{figure}\\
illustrate which signals are needed to be sent from the physical domain to the digital domain in the cloud. Also, in all of these architectures, we assume that in the virtual domain we have a model of the real system that is obtained by system identification algorithms.

\subsection{Architecture I}\label{AA}
Architecture I is illustrated in Fig.~\ref{fig:fig2}. The motivation behind this architecture is that generally, this is the simplest way to make digital twins that may come to mind, and our objective is to investigate the performance of this simple architecture.
As can be seen in Fig.~\ref{fig:fig2}, in this architecture, in the virtual domain we have the model of the real system. Also, we consider the controller in the virtual domain to be exactly the same controller as for the real system in the physical domain. If we use the output signal of the modeled system in the virtual domain as the feedback signal for the virtual controller, in this case, the digital twin and the physical twin will work independently and there is no synchronization between them. Hence, in this architecture, the output signal of the real system, which is measured by sensors, will be sent to the virtual domain via the network and will be used as the feedback signal for the virtual controller. However, since models of systems usually do not work exactly the same as to real systems, using the output signal of the real system as a feedback signal for the controller in the digital domain may not only cause the digital twin not to follow the physical system but the digital twin may also become unstable. We will investigate this in Section V of this paper.
\subsection{Architecture II}
\begin{figure}[htp]
    \centering
    \includegraphics[width=7.5cm, height=3cm]{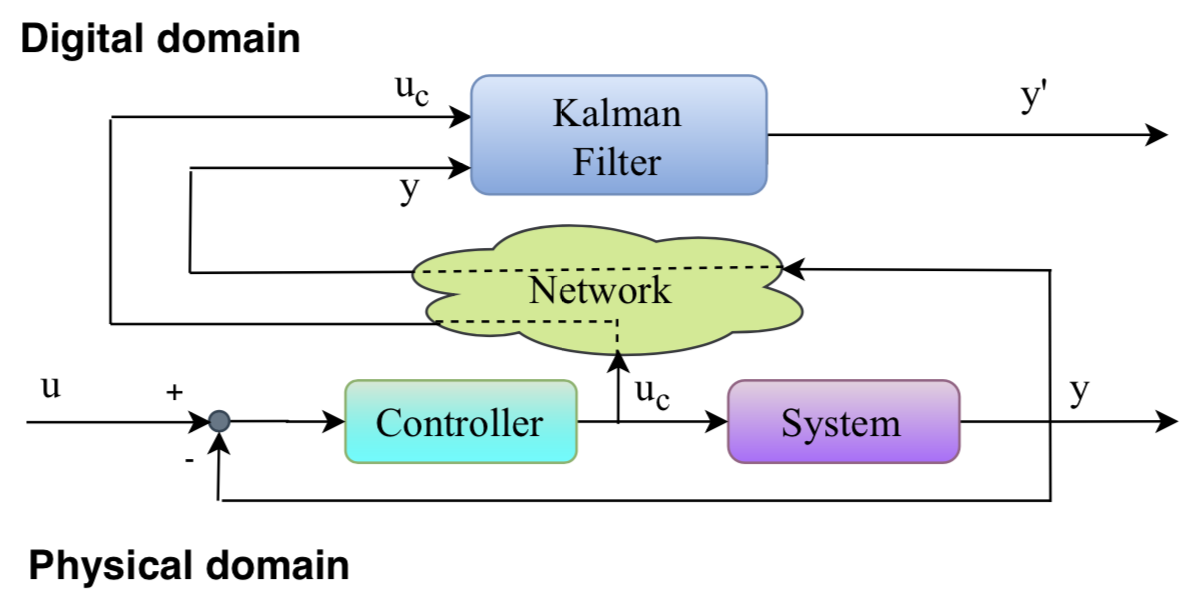}
    \caption{Architecture II}
    \label{fig:fig3}
\end{figure}

Fig.~\ref{fig:fig3} shows architecture II. In this architecture, in order to have the equivalent behavior of the physical system in the digital domain, we try to design an observer. We were motivated to propose this architecture since from a control
theory perspective, we need a kind of observation for making digital twins.
The observer can estimate the physical system’s behavior using input and output signals of the physical system, which are sent through the network to the digital domain. Here, we propose using a Kalman filter as an observer. Because Kalman filter can estimate the actual signal by optimally removing noise from the its input signals. We suppose the state-space model of the physical system is as follows:
\begin{equation}
\begin{array}{ll}
x_{\mathrm{k}+1}=A x_{\mathrm{k}}+B u_{\mathrm{k}}+G w_{\mathrm{k}} & w \rightarrow N(0, Q) \\
y_{\mathrm{k}}=C x_{\mathrm{k}}+F v_{\mathrm{k}} & v \rightarrow N(0, R)
\end{array}
\end{equation}
In this model \(x\) is the state vector, \(y\) is the output signal, \(u\) is the input signal, \(w\) is process noise and \(v\) is measurement noise. Here we consider process noise and measurement noise to be white noise with covariances \(Q\) and \(R\) respectively. 
A Kalman filter for this system will be designed using the following algorithm, which consists of two parts: time update and measurement update\cite{simon2006optimal}. The time update part consists of the following steps: 
\begin{equation}
\text { 1) } \quad \hat{x}_{\mathrm{k} | \mathrm{k}-1}=A_{\mathrm{k}} \hat{x}_{\mathrm{k}-1 | \mathrm{k}-1}+B_{\mathrm{k}} u_{\mathrm{k}}
\end{equation}
\begin{equation}
\text { 2) } P_{\mathrm{k} | \mathrm{k}-1}=G_{\mathrm{k}-1} Q_{\mathrm{k}-1} G_{\mathrm{k}-1}^{\mathrm{T}}+A_{\mathrm{k}-1} P_{\mathrm{k}-1 | \mathrm{k}-1} A_{\mathrm{k}-1}^{\mathrm{T}}
\end{equation}
and the measurement update part consists of following steps: \\
\begin{equation}
\text { 3) } K_{\mathrm{k}}=P_{\mathrm{k} | \mathrm{k}-1} C_{\mathrm{k}}^{\mathrm{T}}\left(C_{\mathrm{k}} P_{\mathrm{k} | \mathrm{k}-1} C_{\mathrm{k}}^{\mathrm{T}}+F_{\mathrm{k}} R_{\mathrm{k}} F_{\mathrm{k}}^{\mathrm{T}}\right)^{-1}
\end{equation}
\begin{equation}
\text { 4) } \hat{x}_{\mathrm{k} | \mathrm{k}}=\hat{x}_{\mathrm{k} | \mathrm{k}-1}+K_{\mathrm{k}}\left(y_{\mathrm{k}}-C_{\mathrm{k}} \hat{x}_{\mathrm{k} | \mathrm{k}-1}\right)
\end{equation}
\begin{equation}
\text { 5) } \quad P_{\mathrm{k} | \mathrm{k}}=\left(I-\mathrm{K}_{\mathrm{k}} C_{\mathrm{k}}\right) P_{\mathrm{k} | \mathrm{k}-1}
\end{equation}

where \(P\) is the estimating covariance matrix and \(K\) is Kalman gain. This observer estimates state variables of the system. The system’s output can also be estimated based on these state variables and using the system model as follows: 
\begin{equation}\label{eq9}
\hat{y}_{\mathrm{k}}=C \hat{x}_{\mathrm{k}}
\end{equation}
so in this way, we can have an estimation of the physical system's behavior in the digital domain.
\subsection{Architecture III}
\begin{figure}[htp]
    \centering
    \includegraphics[width=7.5cm, height=3cm]{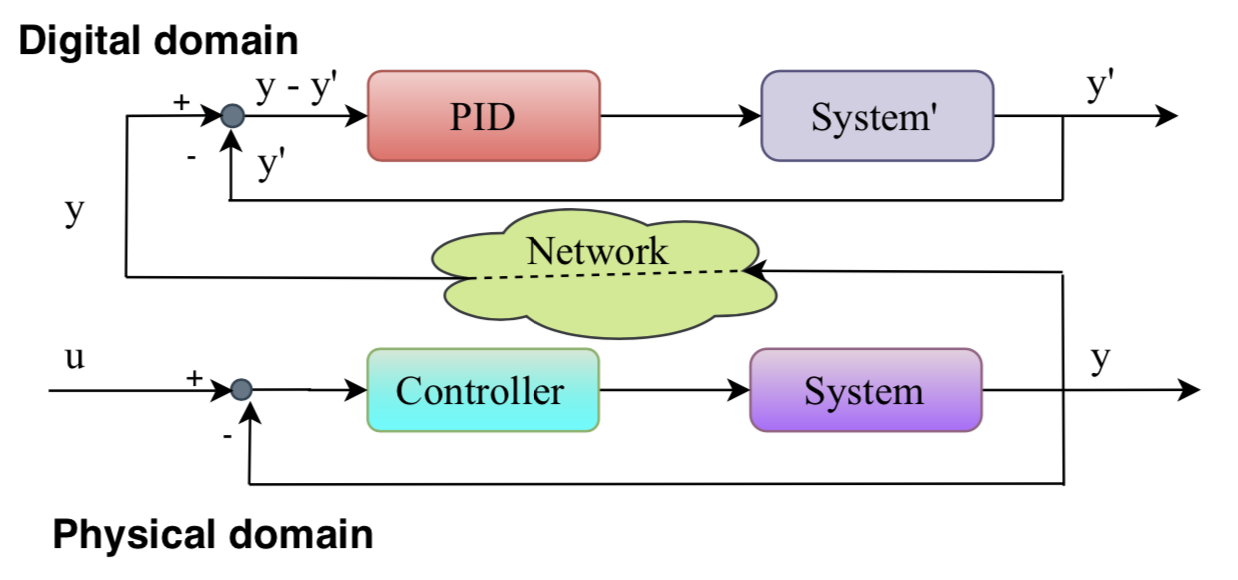}
    \caption{Architecture III}
    \label{fig:fig4}
\end{figure}

Architecture III is demonstrated in Fig.~\ref{fig:fig4}. In this architecture, similar to Architecture I, we have the model of the real system. However, unlike Architecture I, the controller in the virtual domain is not similar to the controller in the physical domain. Here, in order to keep the digital twin in sync with the physical system, we design a new controller for
the digital domain. The objective of designing this controller is controlling the system in the digital domain so that it imitates its physical counterpart’s behavior. For this, we design a PID (Proportional, Integral, Derivative) controller that receives the real system’s output signal as a reference signal. First, by comparing this signal with the output signal of the digital twin an error signal is calculated:
\begin{equation}
e=y_{i}-y'_{i}\label{eq1}
\end{equation}
Then a PID controller is designed as follows:

\begin{equation} \label{eq2}
u'_{c}=K_{p} e+K_{d} \frac{d e}{d t}+K_{i} \int_{0}^{t} e(t) d t
\end{equation}
where \(k_{p}\) is proportional gain, \(k_{d}\) is derivative gain and \(k_{i}\) is integral gain. This controller tries to calculate the control signal so that it minimizes the error between the output of the digital and physical twins. We will evaluate the performance of this architecture in Section V.

\section{Experiment}
We evaluate our proposed architectures using a simulated ball and beam process in Matlab. As the physical twin, we simulate the ball and beam process using the Lagrangian equation of motion for the ball:

\begin{equation}\label{eq10}
0=\left(\frac{J}{R^{2}}+m\right) \ddot{r}+m g \sin \alpha-m r \dot{\alpha}^{2}
\end{equation}

\begin{equation}\label{eq11}
\alpha=\frac{d}{L} \theta
\end{equation}
and we choose its parameters based on \cite{ballandbeam}.

We use a set of input and output data of this system to extract a model based on system identification algorithms, and we use this model as the counterpart of the physical system in the digital domain. In order to simulate the network that is between the physical domain and digital part (cloud), we utilize TrueTime\cite{cervin2003does}, a Simulink toolbox for simulating network transmissions. We consider this network to be Ethernet with \(2.5\)\% packet loss probability and \(40\) ms network delay. Also, we consider process noise and measurement noise as white noise with the covariances \(5\times10^{-6}\) and \(1\times10^{-3}\) respectively. 

The performance of each of the proposed architectures is evaluated by measuring the error between the output signals of the architecture and the physical system \(|y-y’|\). We also measure settling time, the time it takes for the error \(|y-y’|\) between the output signal of the digital twin \(y’\) and the output signal of the physical system y to fall to within \(2\)\% of \(y\).

\section{Results}
\begin{figure}[htp]
    \centering
    \includegraphics[width=8cm]{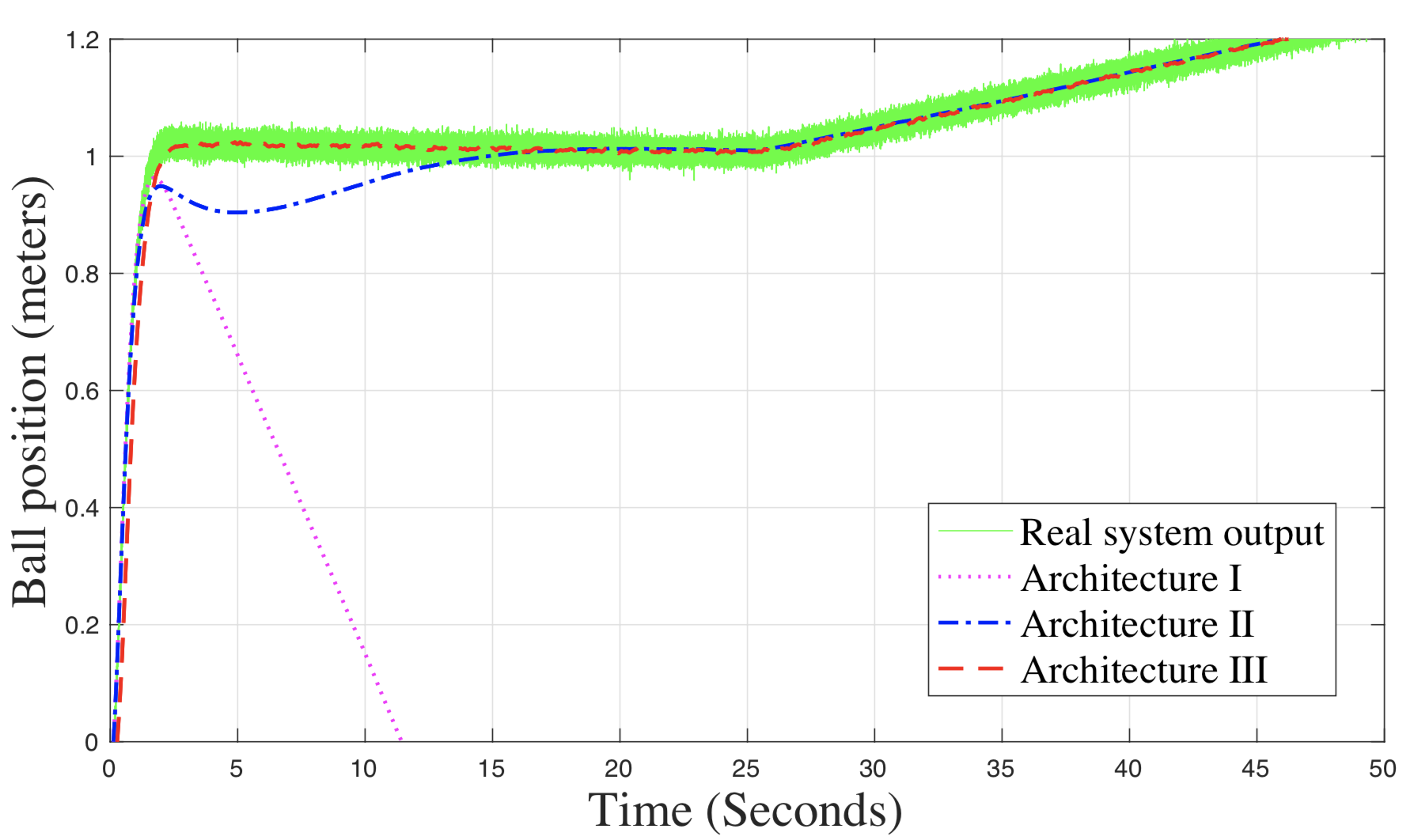}
    \caption{Performance of proposed architectures}
    \label{fig:fig5}
\end{figure}

In this section, results are presented for all three proposed architectures for digital twins. In order to investigate how these proposed architectures are able to follow the physical system’s behavior even when some unexpected changes happen in the system, first, we choose \(1\) as a setpoint for the ball position in the physical domain and then at \(25\) s we add a ramp signal with slope=\(0.01\) to the prior setpoint. As can be seen in Fig.~\ref{fig:fig5},
the digital twin of the ball and beam system made based on Architecture I, not only does not follow its counterpart’s behavior, but also is unstable. This happens since simulated models of systems usually do not work exactly the same as real systems, and so using the output signal of the real system as a feedback signal for the controller in the digital domain causes instability.

\begin{figure}[htp]
    \centering
    \includegraphics[width=7cm, height=5cm]{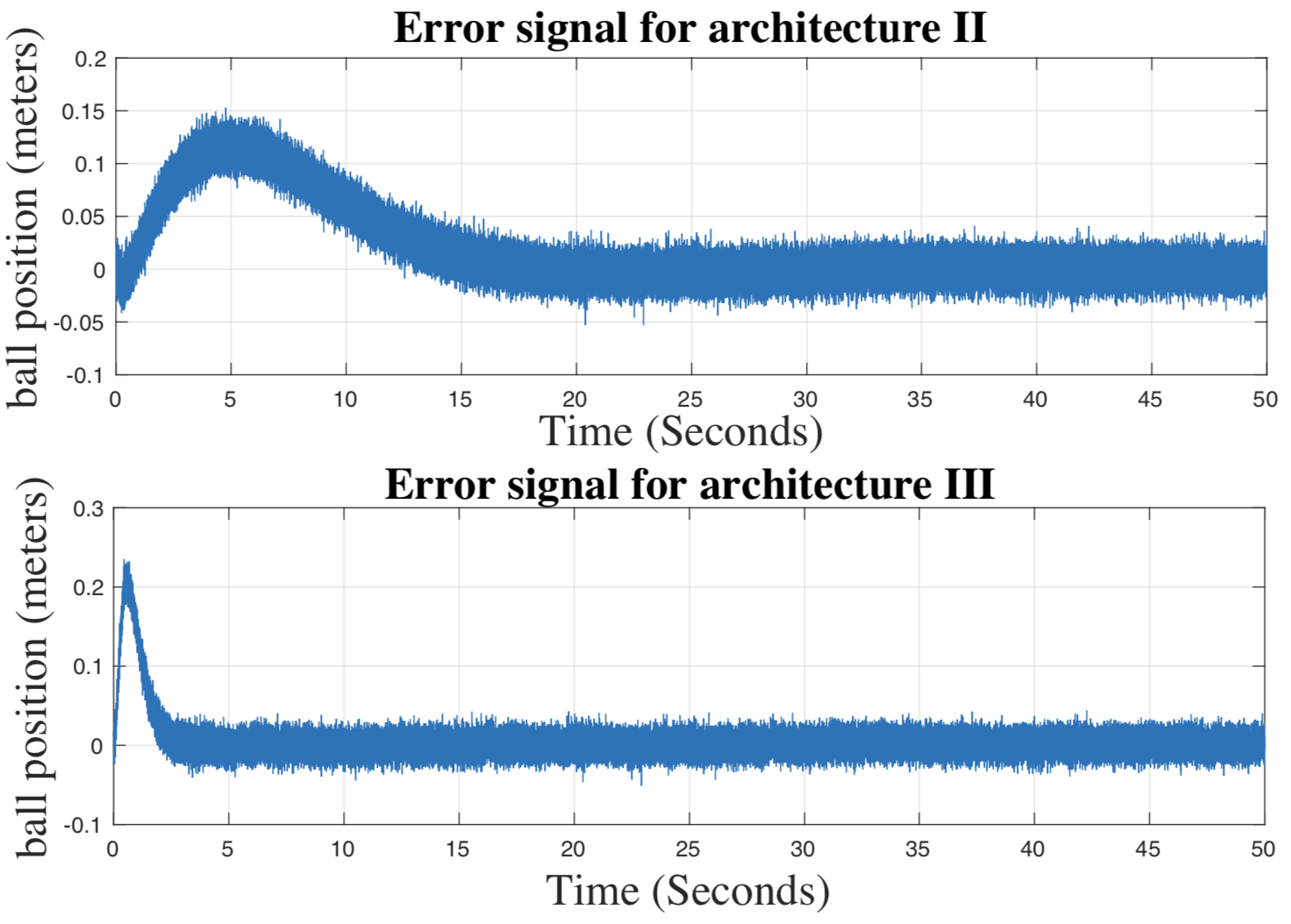}
    \caption{Error signals}
    \label{fig:fig6}
\end{figure}

The dash-dotted curve in Fig.~\ref{fig:fig5} shows the output signal of the digital twin made using Architecture II. As can be seen, in transient state, the digital twin does not behave similar to the physical system, but in steady state, it can follow the physical system’s behavior well. The error signal for this architecture is illustrated in Fig.~\ref{fig:fig6}, and the average of this signal (mean error) equals \(0.0193\). Also, as can be seen in Fig.~\ref{fig:fig5} and~\ref{fig:fig6}, the settling time for architecture II equals \(11.22\) s.

The output signal of the digital twin made using Architecture III is illustrated by the dashed curve in Fig.~\ref{fig:fig5}. This signal follows the position of the ball in the physical domain well. The error signal for this architecture is shown in Fig.~\ref{fig:fig6}, and the average of this signal (mean error) equals \(0.0054\), which is very small. Also, as can be seen in Fig.~\ref{fig:fig5} and~\ref{fig:fig6}, the settling time for Architecture III in this experiment equals \(1.77\) s, shorter than the settling time of Architecture II

As a result, we can say Architecture III has the best performance because it has the smallest mean error and also the fastest settling time. After that, as a second option, Architecture II could be useful if the transient state is not important. However, Architecture I is unstable and is not a good option for making digital twins.

\section{Conclusion}
Digital  Twin is one of the main concepts associated with smart manufacturing. With digital twins, we have virtual replicas of physical systems that precisely mirror the internal behavior of the physical systems. Hence, synchronization to keep the states of digital twins in sync with those of their physical counterparts is crucial. In this paper, we proposed three architectures to make a digital twin that works in sync with its physical counterpart. We evaluated these architectures and we demonstrated the output of Architecture III has the 
smallest error compared with the physical system’s output and also has the smallest settling time.

\section*{Acknowledgment}
This paper was supported by the Celtic-Next project 5G PERFECTA funded by Vinnova, and the SSF project SEC4FACTORY under grant no. SSF RIT17-0032. The authors are part of the Excellence Center at Linköping-Lund on Information Technology (ELLIIT) strategic research area, and the Nordic University Hub on Industrial IoT (HI2OT) funded by NordForsk. Maria Kihl is partially funded by the Wallenberg AI, Autonomous Systems and Software Program (WASP) funded by the Knut and Alice Wallenberg Foundation.

\bibliographystyle{./bibliography/IEEEtran}
\bibliography{conference_041818}

\vspace{12pt}

\end{document}